\documentclass[12pt]{iopart}
\usepackage{iopams}
\usepackage{epsf}
\begin{document}

\title[Nanoscale phase separation in manganites]{Nanoscale phase
separation in manganites}

\author{M Yu Kagan\dag, A V Klaptsov\dag, I V Brodsky\dag,
K I Kugel\ddag, A~O~Sboychakov\ddag~and A L Rakhmanov\ddag}

\address{\dag Kapitza Institute for Physical Problems, Kosygina str. 2,
Moscow, 119334, Russia}

\address{\ddag Institute for Theoretical and Applied Electrodynamics,
Izhorskaya str. 13/19, Moscow, 125412, Russia}
\ead{kagan@kapitza.ras.ru}

\begin{abstract}

     We study the possibility of nanoscale phase separation in
manganites in the framework of the  double exchange model. The
homogeneous canted state of this model is proved to be unstable
toward the formation of small ferromagnetic droplets inside an
antiferromagnetic insulating matrix. For the ferromagnetic
polaronic state we analyze the quantum effects related to the
tails of electronic wave function and a possibility of electron
hopping in the antiferromagnetic background. We find that these
effects lead to the formation of the threshold for the polaronic
state.

\end{abstract}




\section{Introduction}

Manganites, the $\mathrm{Mn}$-based magnetic oxide materials such
as $\mathrm{LaMnO}_3$, have been known for more than 50 years.
Jonker and
 van Santen \cite{Jonker}, and in more details Wollan and Koehler
\cite{Wollan} investigated a rich magnetic structure of
$\mathrm{Ca}$-doped $\mathrm{La}_{1-x}\mathrm{Ca}_x\mathrm{MnO}_3$
and conductivity in these materials. Specifically, a  strong
correlation between magnetic and transport properties in
manganites was observed. These materials exhibit a metal-like
resistance in a ferromagnetic phase and an insulating behavior in
an antiferromagnetic phase.

While the variety of magnetic structures in manganites was
explained by Goodenough \cite{Goodenough} based on the theory of
semicovalent exchange, the correlation between transport and
magnetic properties was for a first time qualitatively explained
by Zener \cite{Zener}. He suggested that the conduction electrons
travel in $\mathrm{La}_{1-x}\mathrm{Ca}_x\mathrm{MnO}_3$ through
the Mn$^{4+}$ ions and each ion carries a fixed magnetic moment
which is strongly ferromagnetically coupled with the spin of a
carrier by a generalized Hund rule. Since a spin of a conduction
electron should be aligned parallel to the local spin, in the
classical picture a conduction electron cannot move in
antiferromagnetically ordered environment. Anderson and Hasegawa
\cite{Anderson} solved a problem of two local spins and one
conduction electron, providing a  strong mathematical background
to Zener ideas. They found that if the bare hopping amplitude was
$t$, than a hopping amplitude of an electron moving between the
two local spins is $t_{eff} = t\cos{\theta}$, where $\theta$ is an
angle between the directions of the local spins. Thus, the kinetic
energy of conduction electrons directly depends on an angle
between the sublattice moments. Moreover, the conduction electrons
tend to align ferromagnetically the local spins, which surround
them. Hence, a competition between ferromagnetic coupling via a
conduction electron (so called double exchange mechanism) and an
antiferromagnetic coupling via superexchange of two neighboring
local spins takes place. De Gennes \cite{deGennes} suggested that
this competition results in a homogeneous canted state, namely an
angle $\theta$ is uniform and monotonically changes from $\pi$
(collinear antiferromagnetic order) to $0$ (collinear
ferromagnetic order) with increasing carriers concentration $x$.
Based on these considerations de Gennes plotted the first phase
diagram of the manganites.

Soon after this work, Nagaev \cite{NagaevQuant,NagaevBig} improved
de Gennes results by considering quantum fluctuations associated
with the local spins. Nagaev proved that electron can move even in
antiferromagnetically ordered phase with a small hopping amplitude
$t/\sqrt{2S+1} $. Nagaev, Kasuya, and Mott also proposed
\cite{NagaevBig,NagaevOne,Kasuya,Mott} that for small electron
concentrations it is more favorable for conduction electrons to
form a self-trapped state (ferromagnetic polaron or ferron) in
antiferromagnetic matrix by creating a small ferromagnetic bubble,
rather than to form a homogeneous canted state in the whole
sample. Thus, it was one of the first hints for nanoscale phase
separation in a double-exchange model made by that time.

Recent growth of interest in manganites was initiated in 1993 by
the discovery of the colossal magnetoresistance (CMR) effect in
doped LaMnO$_3$. The CMR phenomena implies a drastic decrease of
resistivity in manganites in the presence of a magnetic field
\cite{vonHelmolt,Jin}. Soon after discovery of CMR in manganites a
phase diagram of $\mathrm{La}_{1-x}\mathrm{Ca}_x\mathrm{MnO}_3$
was revised \cite{Schiffer}. It was found that a variety of phases
appears in manganites in addition to those predicted by de Gennes.

There are plenty of experimental and theoretical studies of
manganites nowadays. They were initiated first of all by the
potential technological applications of colossal magnetoresistance
phenomena and also by the interesting physics of strong
correlations, which manifests itself in these materials. In
particular, the interaction of spin, charge, and orbital degrees
of freedom in manganites as well as their rich phase diagrams draw
much attention of theorists and experimentalists in recent years.

The important question that has to be answered is about a leading
mechanism responsible for CMR in the optimum doping region
($x\simeq 0.3$). Some authors argue that CMR could be explained in
framework of the double-exchange mechanism alone \cite{Furukawa},
others \cite{Millis} claim that it is necessary to take into
account a lattice interaction (Jahn-Teller polarons), some insist
that percolation-type arguments could explain CMR \cite{Gorkov}.
However, as it was pointed out by Dagotto \etal \cite{Dagotto} and
Arovas and Guinea \cite{Arovas} both analytical and numerical
calculations in various models related to manganites exhibit a
strong tendency toward phase separation in the wide range of
temperatures and concentrations. Thus, it is believed that CMR
phenomena could be understood as a competition and co-existence of
different phases in manganites as well as a phenomena related to
the proximity of the optimum doping region to various
phase-separation thresholds. Note that at higher concentrations
close to half-filling there appears another threshold of phase
separation in the system corresponding again to the formation of
ferromagnetic droplets, but now in a charged ordered insulating
matrix \cite{KagKugKhJETP}.

One of the authors \cite{Kagan} demonstrated that the
double-exchange model at low doping is unstable toward phase
separation and the energy of a homogeneous canted state is higher
than an energy of a self-trapped state corresponding to a
ferromagnetic polaron. Hence a legitimate question arises whether
the stability of a polaronic state is preserved when quantum
fluctuations of spins and tails of wave function of conduction
electrons are taken into account.

\section{Basic theoretical model}

The general chemical formula of the most popular class of
manganites is $\mathrm{Ln}_{1-x}\mathrm{A}_{x}\mathrm{MnO}_3$,
where Ln is a trivalent cation Ln$^{3+}$ (La, Pr, Nb, Sm,
$\ldots$), and A$^{2+}$ is a divalent cation (Ca, Sr, Ba, Mg,
$\ldots$). Oxygen is in a O$^{2-}$ state, and the relative
fraction of Mn$^{4+}$ and Mn$^{3+}$ is controlled by a chemical
doping $x$. This class of manganites has the perovskite structure.
In the cubic lattice environment, the five-fold degenerate
$3d$-orbitals of Mn-ions are split into three lower-energy levels
($d_{xy}$, $d_{yz}$, and $d_{zx}$), usually referred to as
$t_{2g}$, and two higher-energy $e_g$ states ($d_{x^2-y^2}$ and
$d_{3z^2-r^2}$). The $t_{2g}$ levels with three electrons form a
state with a local spin $S=3/2$, whereas delocalized $e_g$ states
contain an electron or are unoccupied depending on the chemical
doping $x$. The $e_g$ states are further split by the static Jahn
- Teller effect and for the simplicity we will treat here only the
lowest $e_g$ state, assuming a Jahn-Teller gap to be large enough
and neglecting any orbital effects in our consideration.

The simplest theoretical model suggested for the explanation of
the properties of manganites is the ferromagnetic Kondo lattice
model ($s-d$ model):
\begin{equation}\label{basic}
    \hat{H}=-J_{H}\sum\limits_{i}\mathbf{S}_i\bsigma_i
    -t\sum\limits_{\langle i,\,j\rangle}Pc^+_{i\sigma}c_{j\sigma}P
    +J_{ff}\sum\limits_{\langle i,\,j\rangle}\mathbf{S}_i\mathbf{S}_j
\end{equation}
The first term in equation (\ref{basic}) represents a strong
on-site Hund's ferromagnetic coupling ($J_H>0$) between the local
spin $S=3/2$ and the spin of a conduction electron. In real
manganites, the Hund's interaction $J_H$ is of the order of 1 eV.
The second term in equation (\ref{basic}) is the kinetic energy of
the conduction electrons. The projection operator $P$ corresponds
to the case of singly occupied $e_g$ orbitals (a strong Hund's
interaction prevents two conduction electrons with different spin
projections from occupying the same site). Note that a strong
electron-lattice interaction significantly reduces the effective
width $W$ of the conduction band ($W=2zt$) resulting in a rather
small hopping amplitude $t\approx 0.3$ eV. The third term in
equation (\ref{basic}) is a weak antiferromagnetic coupling
between local spins on neighboring sites, with $J_{ff}\sim 0.001$
eV. In equation (\ref{basic}), symbols $\langle i,j\rangle$ mean
the summation over $z$ nearest neighbor sites.

In the case of a strong on-site Hund's coupling ($J_H \gg W \gg
J_{ff}$) the model described by the first two terms of the
Hamiltonian (\ref{basic}) is referred to as the double exchange
model. Note that if all local spins are ferromagnetically aligned,
the conduction electrons will move freely in their surrounding.
Thus, model (\ref{basic}) describes the competition between the
direct antiferromagnetic coupling of local spins and the double
exchange via conduction electrons, which tends to order local
spins ferromagnetically. In the strong-coupling limit, Hamiltonian
(\ref{basic}) can be simplified:
\begin{equation}\label{simplified}
    \hat{H}=-\sum\limits_{\langle i,\,j\rangle}t(\theta_{ij})a^+_{i}a_{j}
    +J_{ff}S^2\sum\limits_{\langle i,\,j\rangle}\cos(\theta_{ij})
\end{equation}
where $a^+_i$ and $a_j$ are creation and annihilation operators of
spinless fermions (conduction electrons whose spins $\sigma$ are
aligned parallel to the local spins), $t(\theta_{ij})$ is an
effective hopping amplitude, and $\theta_{ij}$ is an angle between
sublattice moments, as we already discussed. The hopping amplitude
in the case of classical spins ($S\gg 1$) reads:
\begin{equation}\label{deGennes}
    t(\theta) = t\cos(\theta/2)
\end{equation}

In Nagaev's quantum approach the local spins at empty sites have
the maximum projection, $+S$, on the magnetization vector of the
corresponding sublattice. At occupied sites, however, local spin
$\mathbf{S}$ and spin $\bsigma$ of a conduction electron form a
state with the total spin $S+1/2$, but with two possible
projections of it $S\pm 1/2$. So there are two effective bands in
quantum canting corresponding to the two different projections of
the total spin. Their bandwidths read \cite{NagaevBig,Kagan}:
\begin{equation}\label{quantum}
    t_{\pm}(\theta)=\frac{t}{2S+1}\left[\sqrt{2S+1+S^2\cos^2(\theta/2)}
    \pm S\cos(\theta/2)\right]
\end{equation}
The quantum hopping amplitude drastically differs from the
classical de Gennes one. In contrast to the de Gennes picture, an
electron can still move through antiferromagnetic matrix creating
a state with $S^{z}_{tot}=S+1/2$ at one site and a state with
$S^{z}_{tot}=S-1/2$ at a neighboring site (a string-like motion
introduced by Zaanen and Ole\'{s} \cite{Zaanen}):
\begin{equation*}
    \left|S^{z}_{tot}=S+\frac{1}{2}\right\rangle\rightarrow
    \left|S^{z}_{tot}=S-\frac{1}{2}\right\rangle\rightarrow
    \left|S^{z}_{tot}=S+\frac{1}{2}\right\rangle\ldots
\end{equation*}
Hence, there are two equal hopping amplitudes in the case of
electron motion through antiferromagnetic background:
$t_{+}=t_{-}=t/\sqrt{2S+1}$. On the other hand, for ferromagnetic
ordering one gets from equation (\ref{quantum}) $t_{+}=t$ and
$t_{-}=t/(2S+1)$. The proportionality of an effective bandwidth in
the case of quantum canting to $1/\sqrt{S}$ is just an implication
of its quantum nature.

\section{Homogeneous canted state}

An energy of the classical de Gennes state with an account taken
for the antiferromagnetic interaction between the local spins
reads:
\begin{equation}\label{energyDeGennes}
    E=-ztx\cos(\theta/2)+\frac{1}{2}zJ_{ff}S^2\cos(\theta),
\end{equation}
where $z$ is the number of nearest neighbors, and $x$ is a
carriers concentration. The first term in this equation is the
gain in the kinetic energy, and the second term is the loss in the
energy of antiferromagnetic interaction between local spins.
Minimization of the energy (\ref{energyDeGennes}) with respect to
the parameter $\cos(\theta/2)$ yields:
\begin{equation}\label{energyDGcanted}
    E = -\frac{zt^2x^2}{4J_{ff}S^2}-zJ_{ff}S^2,\qquad
    \cos(\theta/2)=\frac{tx}{2J_{ff}S^2}
\end{equation}
Thus, we have a transition from a collinear antiferromagnetic
state for $x=0$ to a collinear ferromagnetic state for
$x=x_{c4}=2J_{ff}S^2/t$. For $0<x<x_{c4}$, the canting angle
($\theta\neq\pi$), and a homogeneous canted state takes place.

Previously, various homogeneous states were considered taking into
account the quantum hopping amplitudes \cite{NagaevBig,Kagan}. In
contrast to the classical case, a collinear antiferromagnetic
state remains energetically favorable up to the critical value of
the carrier concentration $x_{c1}$, which is given by:
\begin{equation}
    x_{c1}=\frac{\pi^4}{3}\left[\frac{8J_{ff}S^2}{zt}\frac{1}{\sqrt{2S+1}}\right]^{3}
\end{equation}
Thus, in a quantum case, the canted state occurs for $x>x_{c1}$,
whereas in the classical case the canted state appears for
arbitrarily low doping levels. At higher doping levels
($x>x_{c1}$), the two-band quantum canted state arises, namely
conduction electrons are in the two bands with total spin
projections $|S^{z}_{tot}=S+1/2\rangle$ and
$|S^{z}_{tot}=S-1/2\rangle$. However at
$x>x_{c2}\approx(27/2)x_{c1}$ the bottom of the second band lies
above the chemical potential level and a one-band state of quantum
canting becomes favorable \cite{Kagan}. Finally, at much higher
carriers concentration ($x>x_{c3}=4J_{ff}S^2/t\sqrt{2S+1}$), a
transition to the classical canted state of de Gennes
(\ref{energyDeGennes}) occurs. Note that for
$x>x_{c4}=2J_{ff}S^2/t$ the canted state transforms into a
collinear ferromagnetic state with the angle $\theta=0$.

To test the stability of the homogeneous state the electronic
compressibility was calculated \cite{Kagan} according to the
standard formula $\kappa^{-1}=d^2E/dx^2$. At very low
concentrations $x<x_{c1}$ the compressibility is positive and a
homogeneous collinear antiferromagnetic state corresponds to at
least a local minimum of energy. However in the range of
concentrations $x_{c1}< x<x_{c4}$ the compressibly $\kappa<0$ and
the canted state is unstable. For example, we can calculate the
compressibility of the classical de Gennes canted state and get:
\begin{equation*}
\kappa^{-1}=-\frac{zt^2}{2J_{ff}S^2}<0.
\end{equation*}
Hence compressibility of this state is negative, i.e.  de Gennes
classical canted state is also unstable. The negative sign of the
compressibility indicates the instability of homogeneous state
toward the phase separation. The simplest case of nanoscale phase
separation corresponds to the formation of small ferromagnetic
polarons inside an antiferromagnetic matrix. This state was
considered in Ref. \cite{Kagan,KagKugelUFN} and it was shown that
a polaronic state is more favorable energetically than all the
homogeneous states in the total range of concentrations $0 < x <
x_{c4}$. Note that magnetic polarons, in this case, correspond to
the electron in the self-trapped ferromagnetic state of a finite
radius inside an antiferromagnetic insulating matrix.

\section{Polaronic state}

As we already discussed, in the case of the classical hopping
amplitude (\ref{deGennes}) a conduction electron may be
self-trapped and form ferromagnetic droplets (magnetic polarons)
inside an antiferromagnetic matrix. The simplest assumption is to
consider that the boundary between the ferromagnetic region and
the antiferromagnetic matrix is abrupt without an extended region
of inhomogeneous canting. Then an energy of a polaronic state
reads:
\begin{equation}\label{energyFerron}
    \fl E=-tx\left(z-\frac{\pi^2a^2}{R^2}\right)+\frac{1}{2}
    zJ_{ff}S^2\,\frac{4\pi}{3}x\left(\frac{R}{a}\right)^3-
    \frac{1}{2}zJ_{ff}S^2\left[1-\frac{4\pi}{3}x\left(\frac{R}{a}\right)^3\right].
\end{equation}
In equation (\ref{energyFerron}), $R$ is a radius of a polaron and
$a$ is a lattice constant. The first term in equation
(\ref{energyFerron}) describes the kinetic energy gain due to the
formation of a ferromagnetic region. The corrections to this term
proportional to $ta^2/R^2$ correspond to the localization energy
of a conduction electron inside a ferromagnetic droplet of radius
$R$. The second term in equation (\ref{energyFerron}) is a loss in
the Heisenberg antiferromagnetic energy of local spins inside the
droplet. Finally, the third term describes the energy of an
antiferromagnetic interaction between local spins in a region
outside the ferromagnetic polarons. The polaron radius is obtained
from the condition of energy minimization $dE/dR=0$. So we have
the following expressions for an energy and a polaron radius:
\begin{eqnarray}\label{energyDGferron}
    E_{pol} = -ztx +\frac{5}{3}\pi^2tx\left(\frac{2zJ_{ff}S^2}{\pi
    t}\right)^{2/5}-\frac{1}{2}zJ_{ff}S^2,\\
    R_{pol}=a\left(\frac{\pi t}{2zJ_{ff}S^2}\right)^{1/5}
\end{eqnarray}
Note that in this case the transition from a polaronic to a
ferromagnetic state occurs when ferromagnetic polarons start to
overlap. The critical concentration for ferromagnetic transition
reads:
\begin{equation}
    x_{c5}=\frac{3}{4\pi}\left(\frac{a}{R}\right)^3=
    \frac{3}{4\pi}\left(\frac{2zJ_{ff}S^2}{\pi t}\right)^{3/5}
\end{equation}

\begin{figure}
\begin{center}
\epsfbox{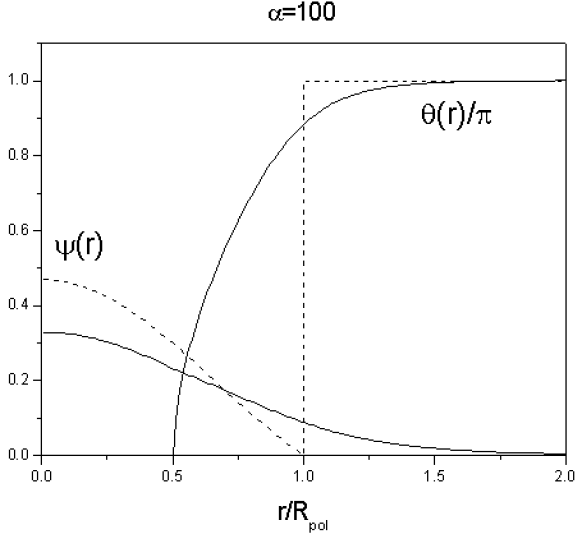}
\end{center}
\caption{\label{wave} The electron wave function $\psi(r)$ and the
canting angle $\theta(r)$ in the quantum case (solid lines) and
the classical case (dashed lines).}
\end{figure}
Now, let us consider the quantum corrections to these results. In
the case corresponding to the quantum hopping amplitude
(\ref{quantum}), a conduction electron can move through
antiferromagnetic background with a heavy mass $m^*\sim \sqrt{S}$,
so it is interesting to study for this case a problem concerning
the stability of a magnetic polaron. Since the analysis of the
discrete model (\ref{simplified}) is rather complicated, we
consider the continuum limit assuming that a radius of a polaron
is much larger than the lattice constant $a$ (further on, we put
$a=1$). A total energy (\ref{simplified}) can now be written in
the following form \cite{Pathak}:
\begin{eqnarray}\label{continuum}
    E=-\int\left[z|\psi|^2+\psi^*\triangle\psi\right]t(\theta)dV+
    zJ_{ff}S^2\int\cos^2(\theta)dV,\\
    t(\theta)=\frac{t}{2S+1}\left[\sqrt{2S+1+S^2\cos^2(\theta/2)}
    + S\cos(\theta/2)\right]
\end{eqnarray}
As one can see from equation (\ref{continuum}), the total energy
will lie between the two limiting  values corresponding
respectively to the motion of the conduction electron via
ferromagnetic or antiferromagnetic backgrounds:
\begin{equation*}
    E_{FM}=-zt<E<E_{AFM}=-\frac{zt}{\sqrt{2S+1}}
\end{equation*}
Since the electron wave function should be normalized
$\int|\psi|^2dV=1$, we minimize the functional
$F=E-t\beta\int|\psi|^2dV$ with respect to parameters $\theta$ and
$\psi$, where $\beta$ is a Lagrange multiplier. The corresponding
Euler-Lagrange equations have the following form:
\begin{eqnarray}\label{equationOne}
    \left[2z\psi+\triangle\psi\right]t(\theta)
    +\triangle\left[t(\theta)\psi\right]-2\beta t\psi=0\\
    \label{equationTwo}
    \left[(z|\psi|^2+\psi^*\triangle\psi)
    \frac{\partial t(\theta)}{\partial\cos(\theta/2)}-
    2zJ_{ff}S^2\cos(\theta/2)\right]\sin(\theta/2)=0
\end{eqnarray}
We solve these two coupled differential equations by the following
iterative procedure \cite{Pathak}: (a) we choose a trial function
for the canting angle $\theta(r)$; (b) we solve the first
differential equation (\ref{equationOne}) to obtain an electron
wave function $\psi(r)$; (c) using an obtained value for $\psi(r)$
we solve equation (\ref{equationTwo}) to get a canting angle
function $\theta(r)$; (d) we return to the step (a) until our
iteration process converges.

\begin{figure}
\begin{center}
\epsfbox{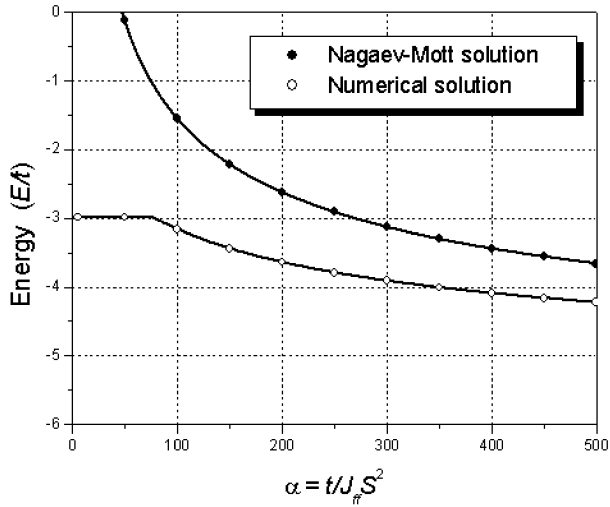}
\end{center}
\caption{\label{Energy} The ground state energy corresponding to
polaron formation in the quantum case (open circles) and the
classical case (solid circles). For local spin $S=3/2$
ferromagnetic and antiferromagnetic states correspond,
respectively, to $E_{FM}=-6t$ and $E_{AFM}=-3t$.}
\end{figure}

Functions $\psi(r)$ and $\theta(r)$ obtained by the numerical
solution of equations (\ref{equationOne}) and (\ref{equationTwo})
are shown in \fref{wave} for a  broad range of the values of
parameter $\alpha=t/J_{ff}S^2$. Note that exact numerical solution
which takes into account both the effects of quantum canting and
tales of the wave function coincides with the classical
Nagaev-Mott solution for $\alpha \rightarrow \infty$. One can see
that the magnetic polaron represents a very good localized object,
and the transition region from the ferromagnetic ordering
($\theta=0$) to an antiferromagnetic matrix ($\theta=\pi$) is
narrow enough. Nevertheless, a polaronic state can disappear at
relatively small value of the parameter $\alpha_c\sim 75$. Indeed,
as one can see from \fref{Energy}, there is a transition from the
polaronic state to a collinear antiferromagnetic state at small
values of parameter $\alpha<\alpha_c$. For this case, the total
energy of a magnetic polaronic state is equal to the bottom of the
band for electron motion through the antiferromagnetic background,
and as a result for $\alpha<\alpha_c$ an electron can move freely
through the antiferromagnetic matrix. Note that to get a more
precise value of $\alpha_c$ we should solve a variational problem
for the functional $F$ on the discrete lattice since for small
values of $\alpha$ a continuous approximation is not accurate
enough. The work along these lines is in progress now.

\section{Conclusion}
The  tendency toward phase-separation is very strong in the
double-exchange model. We have shown that in the wide range of
parameter $\alpha=t/J_{ff}S^2$ a conduction electron forms a
self-trapped state. In this state, an electron is localized in the
ferromagnetic droplet of finite radius embedded in the
antiferromagnetic matrix. This construction seems rather natural
in the de Gennes classical approximation of the double exchange,
where hopping amplitude $t_{eff}= t \cos(\theta/2)$ and electron
\textit{cannot} move through an antiferromagnetic background since
$t_{eff}=0$ for $\theta=\pi$. However, we have proved that even in
the quantum case, when a conduction electron \textit{can} travel
slowly through antiferromagnetic matrix (since $t_{eff}= t/
\sqrt{2S+1}$ for $\theta=\pi$), the polaronic state remains well
defined and stable. Our approach to the one electron problem in an
antiferromagnetic matrix corresponds to a very small doping region
in real materials, where the concentration of charge carriers
(e.g. holes in La$_x$Ca$_{1-x}$MnO$_3$) is low. However we believe
that even for higher concentrations the polaronic picture remains
qualitatively correct. Moreover, at very low concentrations
magnetic polarons should be localized on impurity sites
\cite{NagaevLoc} whereas at concentrations higher than the Mott
threshold they are depinned from the impurities. Preliminary
estimates show that the Mott threshold in manganites corresponds
to $x_M\sim 0.01-0.04$, which is significantly lower than the
critical concentration $x_{c5}\approx0.15$ for the overlap of
magnetic polarons.

Our model of polarons embedded in ferromagnetic matrix allows one
also to calculate a magnetoresistance and a noise spectrum of
manganites, if we suggest that an electron transport takes place
via the hopping of electrons from one polaron to another, while a
polaron itself is immobile. These calculations were carried out in
Ref. \cite{KagRakhKugPRB,Brodsky} and found experimental support
in the recent paper of Babushkina \etal \cite{BabushkinaJPCM}.

If we proceed now to the experimental confirmation of small-scale
phase separated picture we should mention that there already exist
many evidences in favor of nanoscale phase separation in low and
moderatly doped manganites. The confirmation of an inhomogeneous
state in manganites comes from the nuclear magnetic resonance
experiments of Allodi \etal \cite{Allodi1,Allodi2}, where the two
different hyperfine lines corresponding to ferromagnetic and
antiferromagnetic regions were observed. Experiments on neutron
scattering of Biotteau \etal \cite{Hennion1} support the idea of
small ferromagnetic droplets embedded in an antiferromagnetic or a
canted matrix. And if we turn to transport properties, a very
natural picture of electron percolation, which is in agreement
with our model, was experimentally confirmed by Babushkina group
in Ref. \cite{Babushkina}.

Thus the experimental and theoretical picture strongly confirms a
phase-separated state in manganites in the region of low doping.
Moreover, we believe that this picture remains qualitatively
correct for the concentrations optimal for the CMR effect in the
high temperature region $T>T_C$ ($T_C$ is a Curie temperature),
where the ferromagnetic fluctuations of the short range (the
temperature polarons) are present \cite{KagKugelUFN,Krivoglaz}.
Hence, a combination of very intuitive picture of polarons and the
ideas of the percolation theory could provide a correct
description for the behavior of manganites in the wide range of
temperatures and carrier concentrations.

\ack
   The authors acknowledge helpful discussions with D.I. Khomskii,
    V.I. Marchenko, I.A. Fomin, B.E. Meierovich, I. Gonz\'{a}lez,
    M. Hennion, and E. Pchelkin. This work was  supported by the Russian
    Foundation for Basic Research (grant 02-02-16708 and 00-15-96570),
    INTAS grant 01-2008, CRDF grant RP2-2355-MO-02  and
    Russian President Program for Science Support (grant 00-15-9694).

\end{document}